\documentclass[12pt]{article}
\usepackage{amssymb}
\begin{document}
\setcounter{page}{1}
\def\theequation{\arabic{section}.\arabic{equation}}
\def\theequation{\thesection.\arabic{equation}}
\setcounter{section}{0}

\title{Polarization of $\Lambda^0$ hyperons as a signature for the
Quark--Gluon Plasma}

\author{A. Ya. Berdnikov, Ya. A. Berdnikov~\thanks{E--mail:
berdnikov@twonet.stu.neva.ru, State Polytechnic University, Department
of Nuclear Physics, 195251 St. Petersburg, Russian Federation} ,
A. N. Ivanov~\thanks{E--mail: ivanov@kph.tuwien.ac.at, Tel.:
+43--1--58801--14261, Fax: +43--1--58801--14299},
V. A. Ivanova~\thanks{State Polytechnic University, Department of
Nuclear Physics, 195251 St. Petersburg, Russian Federation} ,\\
V. F. Kosmach~\thanks{State Polytechnic University, Department of
Nuclear Physics, 195251 St. Petersburg, Russian Federation} ,
V. M. Samsonov~\thanks{E--mail: samsonov@hep486.pnpi.spb.ru,
St.Petersburg Institute for Nuclear Research, Gatchina, Russian
Federation} , N. I. Troitskaya~\thanks{Permanent Address: State
Polytechnic University, Department of Nuclear Physics, 195251
St. Petersburg, Russian Federation}, V. Thatar
Vento~\thanks{E--mail:thatare@kph.tuwien.ac.at, Permanent Address:
State Polytechnic University, Department of Nuclear Physics, 195251
St. Petersburg, Russian Federation}}

\date{\today}

\maketitle

\begin{center}
{\it Atominstitut der \"Osterreichischen Universit\"aten,
Arbeitsbereich Kernphysik und Nukleare Astrophysik, Technische
Universit\"at Wien, \\ Wiedner Hauptstr. 8-10, A-1040 Wien,
\"Osterreich }
\end{center}

\begin{center}
\begin{abstract}
The momentum distribution of $\Lambda^0$ hyperons produced from the
quark--gluon plasma (QGP) in ultra--relativistic heavy--ion collisions
is calculated in dependence on their polarization. The momentum
distribution of $\Lambda^0$ hyperons is defined by matrix elements of
relativistic quark Wigner operators, which are calculated within the
Effective quark model with chiral $U(3)\times U(3)$ symmetry and the
Quark--Gluon transport theory. We show that the polarization of the
$\Lambda^0$ hyperon depends of the spin of the strange quark that
agrees well with the DeGrand--Miettinen model. We show that
$\Lambda^0$ hyperons, produced from the QGP, are fully unpolarized.
This means that a detection of unpolarized $\Lambda^0$ hyperons,
produced in ultra--relativistic heavy--ion collisions, should serve as
one of the signatures for the existence of the QGP in intermediate
states of ultra--relativistic heavy--ion collisions.\\
\vspace{0.2in}
\end{abstract}
\end{center}

\newpage

\section{Introduction}
\setcounter{equation}{0}

Experimentally a strong polarization of single and
double strange baryons produced in high--energy nuclear reactions has
been well established \cite{GB76}. According to experimental studies
of the polarization of the $\Lambda^0$ hyperon in high--energy nuclear
reactions, summarized by F${\acute{\rm e}}$lix \cite{JF97}, the
polarization of the $\Lambda^0$ hyperon depends on $p_{\perp}$, the
momentum of the $\Lambda^0$ hyperon transversal to the direction of
the incoming beam of colliding nuclei, and $x_f = 2p_z/\sqrt{s}$,
where $p_z$ is the momentum of the $\Lambda^0$ hyperon longitudinal to
the direction of the incoming beam of colliding nuclei and $s$ is the
squared total center--of--mass of event energy \cite{JF97}. Then, it
turned out that the polarization of the $\Lambda^0$ hyperon, produced
in high--energy nucleus--nucleus reactions, is the same for different
targets and different materials with different atomic numbers
\cite{JF97}. Different theoretical models for the description of the
polarization of $\Lambda^0$ hyperons, produced in high--energy nuclear
reactions, have been discussed in details by F${\acute{\rm e}}$lix
\cite{JF99}.

Such a high level of polarization of $\Lambda^0$ hyperons, produced in
high--energy nuclear reactions, can be treated as test for the
classification of nuclear matter in the normal state. Any violation of
high polarization of the $\Lambda^0$ hyperons should testify an
``anomalous'' state of nuclear matter. For the first time the idea
that phase transitions, peculiar for a subnuclear matter in the form
of a quark--gluon plasma (QGP), can provide a symmetry breakdown
leading to the complete depolarization of $\Lambda^0$ hyperons, has
been suggested by Stock \cite{RS82}.

A vanishing polarization of $\Lambda^0$ hyperons as possible signature
of a QGP has been analysed theoretically in
Refs.\cite{AP86}--\cite{MA01} in the semi--classical di--quark,
$s$--quark recombination model \cite{AP86}, Regge--type models
\cite{MA01}. However, as has been pointed out by Bellwied \cite{RB02}
that (i) recent measurements with polarized beams \cite{AB97} seem to
indicate that the semi--classical di--quark, $s$--quark recombination
model is inconsistent with the measured forward--backward asymmetry
($A_N$) and spin transfer asymmetry ($D_{NN}$) and (ii) Regge--type
models and the quark fragmentation model do not necessary lead to the
complete depolarization of $\Lambda^0$ hyperons produced from a QGP.

According to Bellwied \cite{RB02} the measurements of the polarization
of the $\Lambda^0$ hyperons in ultra--relativistic heavy--ion
collisions would be of interest regardless the obtained results, since
nowadays the question of the depolarization of the $\Lambda^0$
hyperon, produced from a QGP in relativistic heavy--ion collisions, is
still open. The measurements of the transversal polarization of
$\Lambda^0$ hyperons produced in 11.6$A\,{\rm GeV/c}$ Au$+$Au
collisions at the Alternating Gradient Synchrotron (AGS) at Brookhaven
National Laboratory carried out by the E896 Collaboration \cite{RB02}
show that for $p_{\perp} \ge 1.5\,{\rm GeV/c}$ and $x_f \ge 0.5$ the
Silicon Drift Detector Array (SDDA) exhibited $P_{\Lambda^0} = (-
25.7\pm 7.5)\%$, whilst the Distributed Drift Chamber (DDC) gave
$P_{\Lambda^0} = (- 23\pm 11)\%$. These results show \cite{RB02} that
in the heaviest collision systems $\Lambda^0$ hyperons are still
polarized at freeze--out. The direction of the $\Lambda^0$ hyperon
spin is slightly affected by the rescattering phase after
hadronization. These measurements are first to confirm the
polarization of $\Lambda^0$ hyperons produced in heavy--ion collisions
\cite{RB02}.

The problem of polarization of $\Lambda^0$ hyperons as a probe for a
QGP has been recently discussed by Ayala {\it et al.}  \cite{AA02}. For
the description of the polarization of the $\Lambda^0$ hyperons for
densities below a critical density for the formation of a QGP the
authors have used the di--quark, $s$--quark recombination mechanism
within the DeGrand--Miettinen model \cite{JF97,TD81}.  Assuming that
for densities higher than the critical density of the formation of a
QGP a recombination mechanism of the $\Lambda^0$ polarization should
be changed to the coalescence of free valence quarks, the authors have
found that the polarization of the $\Lambda^0$ hyperons should depend
on the relative contribution of each process to the total number of
the $\Lambda^0$ hyperons produced in the collision.

In this paper we suggest to analyse the problem of the polarization of
$\Lambda^0$ hyperons, produced from a QGP in ultra-relativistic
heavy--ion collisions, in terms of the momentum distribution of the
number of $\Lambda^0$ hyperons $N_{\Lambda^0}(\vec{k}\,)$ defined by
\cite{FC74,UH1}
\begin{eqnarray}\label{label1.1}
E_{\vec{k}}\,\frac{d^3N_{\Lambda^0}}{d^3k} =
\frac{2}{(2\pi)^3}\int_{\Sigma_f} \theta(k^0)\,f_{\Lambda^0}(x,k)
k^{\mu}d\sigma_{\mu}(x),
\end{eqnarray}
where the common factor 2 is the spin degeneracy of the
$\Lambda^0$ hyperon and $f_{\Lambda^0}(x,k)$ is its distribution
function, and $d\sigma_{\mu}(x)$ is a normal vector to the freeze--out
3--dimensional surface $\Sigma_f$ in the configuration space--time
\cite{FC74,UH1}. The distribution function $f_{\Lambda^0}(x,k)$ is
taken in the J\"uttner form \cite{FJ28,SG1} (see also
\cite{FC74,UH1})
\begin{eqnarray}\label{label1.2}
f_{\Lambda^0}(x,k) = \frac{1}{\displaystyle e^{\textstyle\, (k\cdot
U(x) - \mu_{\Lambda^0}(x)/T(x)} + 1},
\end{eqnarray}
where $U(x)$, $\mu_{\Lambda^0}(x)$ and $T(x)$ are the 4--dimensional
hydrodynamical local velocity, the baryon chemical potential and the
temperature, and $k^{\mu} = (k^0, \vec{k}\,)$ with $k^0 = E =
\sqrt{\vec{k}^{\,2} + m^2_{\Lambda^0}}$ is the 4--momentum of the
$\Lambda^0$ hyperon, $\theta(k^0)$ is the Heaviside function.

In order to take into account polarization properties of the
$\Lambda^0$ hyperon it is convenient to determine the distribution
function $f_{\Lambda^0}(x,k)$ in terms of the positive energy Wigner
function $W^{(+)}_{\Lambda^0}(x,k)_{\beta\alpha}$
\cite{SG1}--\cite{UH2}
\begin{eqnarray}\label{label1.3}
\theta(k^0)f_{\Lambda^0}(x,k) = \sum_{\alpha}
W^{(+)}_{\Lambda^0}(x,k)_{\alpha\alpha} = {\rm
tr}W^{(+)}_{\Lambda^0}(x,k),
\end{eqnarray}
where $W^{(+)}_{\Lambda^0}(x,k)_{\beta\alpha}$ is defined by
\cite{SG1}--\cite{UH2}
\begin{eqnarray}\label{label1.4}
W^{(+)}_{\Lambda^0}(x,k)_{\beta\alpha} = \int
\frac{d^4y}{(2\pi)^4}\,\theta(k^0)\,e^{\textstyle\,-ik\cdot
y}\Big\langle \Omega\Big|:\bar{\psi}_{\Lambda^0}\Big(x +
\frac{1}{2}\,y\Big)_{\beta}\psi_{\Lambda^0}\Big(x -
\frac{1}{2}\,y\Big)_{\alpha}:\Big|\Omega\Big\rangle.
\end{eqnarray}
Here $|\Omega\rangle$ is the vacuum wave function,
$\psi_{\Lambda^0}(z)$ is the interpolating field for free
$\Lambda^0$ hyperons, and $:\ldots:$ indicates the normal ordering
\cite{SG1}.

The interpolating field $\psi_{\Lambda^0}(x \pm y/2)$ we represent as
\begin{eqnarray}\label{label1.5}
\psi_{\Lambda^0}\Big(x \pm \frac{1}{2}\,y\,\Big) &=&
\frac{1}{(2\pi)^3}\sum_{\sigma = \pm
1/2}\sqrt{\frac{m_{\Lambda^0}}{2}}\int
\frac{d^3p}{E(\vec{p}\,)}\,\Big[u(\vec{p},\sigma)
a_{\Lambda^0}(\vec{p},\sigma)\,e^{\textstyle\,-ip\cdot(x \pm
y/2)}\nonumber\\ && + v(\vec{p},\sigma) b^{\dagger}_{\bar{\Lambda}^0}(
\vec{p},\sigma)\,e^{\textstyle\,-ip\cdot(x \pm y/2)},
\end{eqnarray}
where $u(\vec{p},\sigma)$ and $v(\vec{p},\sigma)$ are the Dirac
bispinors normalized by
\begin{eqnarray}\label{label1.6}
\bar{u}(\vec{p},\sigma)u(\vec{p},\sigma\,'\,) = -
\bar{v}(\vec{p},\sigma) v(\vec{p},\sigma\,'\,) =
\delta_{\sigma\sigma\,'}.
\end{eqnarray}
The operators $a_{\Lambda^0}(\vec{p},\sigma)$ and
$a^{\dagger}_{\Lambda^0}(\vec{p},\sigma)$ annihilate and create the
$\Lambda^0$ hyperon with quantum numbers $(E,\vec{p},\sigma)$,
whereas the operators $b_{\bar{\Lambda}^0}(\vec{p},\sigma)$ and
$b^{\dagger}_{\bar{\Lambda}^0}(\vec{p},\sigma)$ annihilate and create
the anti--$\Lambda^0$ hyperon ($\bar{\Lambda}^0$--hyperon) with
quantum numbers $(E, \vec{p},\sigma)$. These operators satisfy
canonical relativistic covariant anticommutation relations \cite{SG1}
\begin{eqnarray}\label{label1.7}
\{a_{\Lambda^0}(\vec{p},\sigma),
a^{\dagger}_{\Lambda^0}(\vec{q},\lambda)\} &=&
\{b_{\Lambda^0}(\vec{p},\sigma),
b^{\dagger}_{\Lambda^0}(\vec{q},\lambda)\} = (2\pi)^3\,
2E(\vec{p}\,)\,\delta^{(3)}(\vec{p} -
\vec{q}\,)\delta_{\sigma\lambda},\nonumber\\
\{a_{\Lambda^0}(\vec{p},\sigma),a_{\Lambda^0}(\vec{q},\lambda)\} &=&
\{b_{\Lambda^0}(\vec{p},\sigma), b_{\Lambda^0}(\vec{q},\lambda)\} = 0.
\end{eqnarray}
The vacuum expectation values of the products of operators of creation
and annihilation of $\Lambda^0$ and $\bar{\Lambda}^0$ hyperons are
equal to \cite{SG1,DT57}
\begin{eqnarray}\label{label1.8}
\langle \Omega|a^{\dagger}_{\Lambda^0}(\vec{q},\lambda)
a_{\Lambda^0}(\vec{p},\sigma)|\Omega\rangle = (2\pi)^3
2E(\vec{p}\,)\,\delta^{(3)}(\vec{q} -
\vec{p}\,)\,f_{\Lambda^0}(0,p)\,\delta_{\lambda\sigma},\nonumber\\
\langle \Omega|b^{\dagger}_{\bar{\Lambda}^0}(\vec{q},\lambda)
b_{\bar{\Lambda}^0}(\vec{p},\sigma)|\Omega\rangle = (2\pi)^3
2E(\vec{p}\,)\,\delta^{(3)}(\vec{q} -
\vec{p}\,)\,f_{\bar{\Lambda}^0}(0,p)\,\delta_{\lambda\sigma}.
\end{eqnarray}
Computing the vacuum expectation value in(\ref{label1.4}) we define
the Wigner function for the polarized $\Lambda^0$ hyperon as
\begin{eqnarray}\label{label1.9}
W^{(+)}_{\Lambda^0}(x,k)_{\beta\alpha} =
\theta(k^0)\,\rho(k,\zeta)_{\alpha\beta}\,f_{\Lambda^0}(0,k).
\end{eqnarray}
Here $\rho(k,\zeta)_{\alpha\beta}$ is the spin density matrix
\begin{eqnarray}\label{label1.10}
\rho(k,\zeta)_{\alpha\beta} = \frac{\hat{k} + m_{\Lambda^0}}{4
m_{\Lambda^0}}\,(1 + \gamma^5\hat{\zeta}),
\end{eqnarray}
where $\hat{a} = \gamma_{\mu}a^{\mu}$ and $\zeta^{\mu}$ is the
polarization vector given by \cite{SG1}
\begin{eqnarray}\label{label1.11}
\zeta^{\mu} =
\Big(\frac{\vec{p}\cdot\vec{\zeta}}{m_{\Lambda^0}},\vec{\zeta} +
\frac{\vec{k}\,(\vec{k}\cdot\vec{\zeta})}{m_{\Lambda^0}(E(\vec{k}\,) +
m_{\Lambda^0})}\,\Big)
\end{eqnarray}  
and $\vec{\zeta}$ is a unit vector, $\vec{\zeta}^{\;2} = 1$.

In the non--relativistic limit $m_{\Lambda^0} \gg |\vec{k}\,|$ the
spin density matrix $\rho(k,\zeta)$ reduces to the standard form
\begin{eqnarray}\label{label1.12}
\hat{\rho}(k,\zeta) = \frac{1}{2}\,(1 + \vec{\sigma}\cdot
\vec{\zeta}\,),
\end{eqnarray}  
where $\vec{\sigma}$ are $2\times 2$ Pauli matrices.

Substituting (\ref{label1.9}) in (\ref{label1.3}) and computing the
trace over Dirac matrices we prove the relation (\ref{label1.3}). One
can replace $f_{\Lambda^0}(0,k)$ by $f_{\Lambda^0}(x,k)$ for the
$\Lambda^0$--hyperon gas in the quasi--equilibrium state
\cite{UH1,SG1}. The result should be valid for any non--equilibrium
state of the $\Lambda^0$--hyperon gas.

Thus, we have shown that the $\Lambda^0$ hyperon, treated as a
point--like particle and produced from the QGP, should be
unpolarized. It is well--known that quark degrees of freedom play an
important role for production and polarization of the $\Lambda^0$
hyperon in hadron--hadron and nucleus--nucleus collisions. Therefore,
the polarization of $\Lambda^0$ hyperons, produced from the QGP,
should be investigated at the quark level.

The work is organized as follows. In Section 2 we define a momentum
distribution of $\Lambda^0$ hyperons, produced from the QGP, in terms
of the matrix elements of the relativistic quark Wigner operators. In
Section 3 we calculate the matrix elements of the relativistic quark
Wigner operators within the Effective quark model with chiral
$U(3)\times U(3)$ symmetry and the Quark--Gluon transport theory. In
Section 4 we analyse the dependence of the momentum distribution of
the number of $\Lambda^0$ hyperons on the polarization of $\Lambda^0$
hyperons. We show that $\Lambda^0$ hyperons produced from the QGP are
unpolarized. In the Conclusion we discuss the obtained results. In
Appendix A and B we calculate the momentum integrals defining the
momentum distribution of $\Lambda^0$ hyperons in our approach.

\section{Momentum distribution of  $\Lambda^0$ hyperons. 
Relativistic quark Wigner operator }
\setcounter{equation}{0}

At the quark level the momentum distribution of the number
$N_{\Lambda^0}(\vec{k}\,)$ of $\Lambda^0$ hyperons, produced from  a QGP
in ultra-relativistic heavy--ion collisions, can be defined by
\cite{UH1}
\begin{eqnarray}\label{label2.1}
\hspace{-0.3in}&&E_{\vec{k}}\,\frac{d^3N_{\Lambda^0}(\vec{k}\,)}{d^3k}
= \nonumber\\\hspace{-0.3in}&&= \sum_{\textstyle \lambda = \pm
1/2}\sum_{\textstyle q =
u,d,s}\int_{\textstyle\Sigma}d\sigma^{\mu}(x)k_{\mu}\,\theta(k^0)\int
d^4p\,\langle
\Lambda^0(k,\lambda)|\hat{W}^{(+)}_q(x,p)|\Lambda^0(k,\lambda)\rangle,
\end{eqnarray}
where $\Sigma$ is a surface of a ``freeze--out'' isotherm \cite{UH1}.
Then, $\hat{W}^{(q)}(x,p)$ is the relativistic $q$--quark Wigner
operator determined by \cite{UH2}
\begin{eqnarray}\label{label2.2}
\hat{W}^{(+)}_q(x,p) = \theta(p^0)\int
\frac{d^4y}{(2\pi)^4}\,e^{\textstyle -ip\cdot y} :\bar{q}(x +
\frac{1}{2}\,y)\otimes q(x - \frac{1}{2}\,y):,
\end{eqnarray}
where dots $:\ldots:$ indicate the normal ordering.

The vacuum expectation value of the relativistic quark Wigner operator
\begin{eqnarray}\label{label2.3}
\langle \Omega|\hat{W}^{(+)}_q(x,p)|\Omega\rangle
\end{eqnarray}
is equal to the distribution function of the $q$--quark
\cite{SG1,UH2}. In turn, the matrix element
\begin{eqnarray}\label{label2.4} 
\langle\Lambda^0(k,\lambda)|\hat{W}^{(+)}_q(x,p)|\Lambda^0(k,\lambda)
\rangle
\end{eqnarray}
describes a projection of the $q$--quark, appearing in the
intermediate state of ultra--relativistic heavy--ion collisions, on
the physical states of $\Lambda^0$--hyperons on--mass shell.

The calculation of the matrix elements we suggest to carry out within
the Effective quark model of baryons with chiral $U(3)\times U(3)$
symmetry \cite{AI99}--\cite{AI01} and the Quark--Gluon transport
theory \cite{UH2}.

\section{Matrix element $\langle\Lambda^0(k,\lambda)|
\hat{W}^{(+)}_q(x,p)|\Lambda^0(k,\lambda)\rangle$}
\setcounter{equation}{0}

In this section calculate the matrix element
(\ref{label2.4}). Using the reduction technique we get
\cite{AI99}--\cite{AI01}
\begin{eqnarray}\label{label3.1} 
\hspace{-0.3in}&&\langle\Lambda^0(k,\lambda)|
\hat{W}^{(+)}_q(x,p)|\Lambda^0(k,\lambda) \rangle = \lim_{\textstyle
k^2\to M^2_{\Lambda^0}} \int d^4x_1 d^4x_2 \,e^{\textstyle
ik\cdot(x_1 - x_2)}\,\bar{u}_{\Lambda^0}(k,\lambda)\nonumber\\
\hspace{-0.3in}&&\overrightarrow{\Bigg(i\gamma^{\nu}
\frac{\partial}{\partial x^{\nu}_1} - M_{\Lambda^0}\Bigg)}\langle
\Omega|{\rm T}(\psi_{\Lambda^0}(x_1)\hat{W}^{(+)}_q(x,p)
\bar{\psi}_{\Lambda^0}(x_2))|\Omega\rangle\overleftarrow{\Bigg(-
i\gamma^{\alpha}\frac{\partial}{\partial x^{\alpha}_2} -
M_{\Lambda^0}\Bigg)}u_{\Lambda^0}(k,\lambda),\nonumber\\
\hspace{-0.3in}&&
\end{eqnarray}
where $\psi_{\Lambda^0}(x)$ and $u_{\Lambda^0}(k,\lambda)$ are the
interpolating field operator and the Dirac bispinor of the $\Lambda^0$
hyperon, respectively.

In order to describe the r.h.s. of Eq.(\ref{label3.1}) at the quark
level we follow \cite{AI99} and use the equations of motion
\begin{eqnarray}\label{label3.2}
\overrightarrow{\Bigg(i\gamma^{\nu}\frac{\partial}{\partial x^{\nu}_1}
- M_{\Lambda^0}\Bigg)}\,\psi_{\Lambda^0}(x_1) &=& \frac{g_{\rm
B}}{\sqrt{2}}\,\eta_{\rm \Lambda^0}(x_1),\nonumber\\
\bar{\psi}_{\Lambda^0}(x_2)\overleftarrow{\Bigg(-
i\gamma^{\alpha}\frac{\partial}{\partial x^{\alpha}_2} -
M_{\Lambda^0}\Bigg)} &=& \frac{g_{\rm
B}}{\sqrt{2}}\,\bar{\eta}_{\Lambda^0}(x_2),
\end{eqnarray}
where $M_{\Lambda^0} = 1116\,{\rm MeV}$ is a mass of the $\Lambda^0$
hyperon. Then, $\eta_{\rm \Lambda^0}(x_1)$ and $\bar{\eta}_{\rm
\Lambda^0}(x_2)$ are the three--quark current densities \cite{AI99}
\begin{eqnarray}\label{label3.3}
\eta_{\rm \Lambda^0}(x_1) &=& -
\sqrt{\frac{2}{3}}\,\varepsilon^{ijk}\,\{[\bar{u^c}_i(x_1)
\gamma^{\mu}s_j(x_1)]\gamma_{\mu}\gamma^5 d_k(x_1) -
[\bar{d^c}_i(x_1)\gamma^{\mu}s_j(x_1)]\gamma_{\mu}\gamma^5
u_k(x_1)\},\nonumber\\ \bar{\eta}_{\rm \Lambda^0}(x_2) &=&+
\sqrt{\frac{2}{3}}\,\varepsilon^{ijk}\{\bar{d}_i(x_1)\gamma^{\mu}
\gamma^5[\bar{s}_j(x_1)\gamma_{\mu}u^c_k(x_1)] -
\bar{u}_i(x_1)\gamma^{\mu}\gamma^5
[\bar{s}_j(x_1)\gamma_{\mu}d^c_k(x_1)]\},
\end{eqnarray}
where $i,j$ and $k$ are colour indices and $\bar{\psi^{\,c}}(x) =
\psi(x)^T C$ and $C = - C^T = - C^{\dagger} = - C^{-1} $ is the
conjugation charge, $T$ is a transposition, and $g_{\rm B}$ is the
phenomenological coupling constant of the low--lying baryon octet
$B_8(x)$ coupled to three--quark current \cite{AI99}
\begin{eqnarray}\label{label3.4}
\hspace{-0.5in}{\cal L}^{(\rm B)}_{\rm int}(x) = \frac{g_{\rm
B}}{\sqrt{2}}\,\bar{B}_8(x)\eta_8(x) + {\rm h.c.}\,,
\end{eqnarray}
where the numerical value of $g_{\rm B}$ is equal to $g_{\rm B}=
1.34\times 10^{-4}\,{\rm MeV}^{-2}$ \cite{AI99}.

Substituting (\ref{label3.2}) in (\ref{label3.1}) we obtain
\begin{eqnarray}\label{label3.5} 
\hspace{-0.5in}\langle\Lambda^0(k,\lambda)|
\hat{W}^{(+)}_q(x,p)|\Lambda^0(k,\lambda) \rangle &=& \frac{1}{2}\,
g^2_{\rm B}\int d^4x_1 d^4x_2 \,e^{\textstyle ik\cdot(x_1 -
x_2)}\,\bar{u}^{(\alpha)}_{\Lambda^0}(k,\lambda)\nonumber\\
\hspace{-0.5in}&\times&{\rm tr}\{\langle \Omega|{\rm
T}(\eta^{(\alpha)}_{\Lambda^0}(x_1)\hat{W}^{(+)}_q(x,p)
\bar{\eta}^{(\beta)}_{\Lambda^0}(x_2))|\Omega\rangle\}
u^{(\beta)}_{\Lambda^0}(k,\lambda),
\end{eqnarray}
where the trace should be calculated over colour and spinorial indices
of quark fields. Recall, that the $\Lambda^0$ hyperon should be kept
on mass--shell $k^2 = M^2_{\Lambda^0}$.

The matrix element ${\rm tr}\{\langle \Omega|{\rm
T}(\eta^{(\alpha)}_{\Lambda^0}(x_1)\hat{W}^{(+)}_q(x,p)
\bar{\eta}^{(\beta)}_{\Lambda^0}(x_2))|\Omega\rangle\}$ is defined by
\begin{eqnarray}\label{label3.6}
&&{\rm tr}\{\langle \Omega|{\rm
T}(\eta^{(\alpha)}_{\Lambda^0}(x_1)\hat{W}^{(+)}_q(x,p)
\bar{\eta}^{(\beta)}_{\Lambda^0}(x_2))|\Omega\rangle\} = \int
\frac{d^4y}{(2\pi)^4}\,\theta(p^0)\,e^{\textstyle -ip\cdot
y}\nonumber\\ &&\times\,{\rm tr}\{\langle \Omega|{\rm
T}(\eta^{(\alpha)}_{\Lambda^0}(x_1):\bar{q}(x + \frac{1}{2}\,y)\otimes
q(x - \frac{1}{2}\,y): \bar{\eta}^{(\beta)}_{\Lambda^0}(x_2))
|\Omega\rangle\}.
\end{eqnarray}
The calculation of the vacuum expectation value in the r.h.s. of
(\ref{label3.6}) we carry out by the example of
$\hat{W}^{(u)}(x,p)$. This calculation runs in the way
\begin{eqnarray}\label{label3.7} 
\hspace{-0.3in}&&{\rm tr}\{\langle \Omega|{\rm
T}(\eta^{(\alpha)}_{\Lambda^0}(x_1):\bar{u}(x + \frac{1}{2}\,y)\otimes
u(x - \frac{1}{2}\,y): \bar{\eta}^{(\beta)}_{\Lambda^0}(x_2)) |\Omega
\rangle\} = -
\frac{2}{3}\,\varepsilon^{ijk}\varepsilon^{i'j'k'}\nonumber\\
\hspace{-0.3in}&&\times\,{\rm tr}\{\langle \Omega|{\rm
T}(\{[\bar{u^c}_i(x_1) \gamma^{\mu}s_j(x_1)](\gamma_{\mu}\gamma^5
d_k(x_1))^{(\alpha)} -
[\bar{d^c}_i(x_1)\gamma^{\mu}s_j(x_1)](\gamma_{\mu}\gamma^5
u_k(x_1))^{(\alpha)}\} \nonumber\\
\hspace{-0.3in}&&\times\,:\bar{u}_{\ell}(x + \frac{1}{2}\,y)\otimes
u_{\ell}(x - \frac{1}{2}\,y):\{(\bar{d}_{i'}(x_2)\gamma^{\nu}
\gamma^5)^{(\beta)}[\bar{s}_{j'}(x_2)\gamma_{\nu}u^c_{k'}(x_2)] -
(\bar{u}_{i'}(x_2)\gamma^{\nu}\gamma^5)^{(\beta)} \nonumber\\
\hspace{-0.3in}&&
\times\,[\bar{s}_{j'}(x_2)\gamma_{\nu}d^c_{k'}(x_2)]\})|\Omega\rangle\}
= - \frac{2}{3}\,\varepsilon^{ijk}\varepsilon^{i'j'k'}{\rm tr}\{\langle
\Omega|{\rm T}([\bar{u^c}_i(x_1) \gamma^{\mu}s_j(x_1)](\gamma_{\mu}\gamma^5
d_k(x_1))^{(\alpha)}\nonumber\\
\hspace{-0.3in}&&\times\,\bar{u^c}_{\ell}(x + \frac{1}{2}\,y)\otimes
u^c_{\ell}(x - \frac{1}{2}\,y)(\bar{d}_{i'}(x_2)\gamma^{\nu}
\gamma^5)^{(\beta)}[\bar{s}_{j'}(x_2)\gamma_{\nu}
u^c_{k'}(x_2)])|\Omega \rangle\} \nonumber\\
\hspace{-0.3in}&& +
\frac{2}{3}\,\varepsilon^{ijk}\varepsilon^{i'j'k'}{\rm tr}\{\langle
0|{\rm T}([\bar{d^c}_i(x_1)\gamma^{\mu}s_j(x_1)]\nonumber\\
\hspace{-0.3in}&&\times\,(\gamma_{\mu}\gamma^5
u_k(x_1))^{(\alpha)}\bar{u}_{\ell}(x - \frac{1}{2}\,y)u_{\ell}(x +
\frac{1}{2}\,y)(\bar{u}_{i'}(x_2)\gamma^{\nu}\gamma^5)^{(\beta)}
[\bar{s}_{j'}(x_2)\gamma_{\nu}d^c_{k'}(x_2)])|\Omega
\rangle\}\nonumber\\
\hspace{-0.3in}&&+
\frac{2}{3}\,\varepsilon^{ijk}\varepsilon^{i'j'k'}{\rm tr}\{\langle
\Omega |{\rm T}([\bar{u^c}_i(x_1)
\gamma^{\mu}s_j(x_1)](\gamma_{\mu}\gamma^5
d_k(x_1))^{(\alpha)}\bar{u}_{\ell}(x - \frac{1}{2}\,y)\otimes
u_{\ell}(x + \frac{1}{2}\,y) \nonumber\\
\hspace{-0.3in}&&\times\,
(\bar{u}_{i'}(x_2)\gamma^{\nu}\gamma^5)^{(\beta)}
[\bar{s}_{j'}(x_2)\gamma_{\nu}d^c_{k'}(x_2)])|\Omega \rangle\}
\nonumber\\
\hspace{-0.3in}&&+
\frac{2}{3}\,\varepsilon^{ijk}\varepsilon^{i'j'k'}{\rm tr}\{\langle
\Omega |{\rm
T}([\bar{d^c}_i(x_1)\gamma^{\mu}s_j(x_1)](\gamma_{\mu}\gamma^5
u_k(x_1))^{(\alpha)}\bar{u}_{\ell}(x - \frac{1}{2}\,y)\otimes
u_{\ell}(x + \frac{1}{2}\,y)\nonumber\\
\hspace{-0.3in}&&\times\,(\bar{d}_{i'}(x_2)\gamma^{\nu}
\gamma^5)^{(\beta)}[\bar{s}_{j'}(x_2) \gamma_{\nu}
u^c_{k'}(x_2)])|\Omega \rangle\}.
\end{eqnarray}
In the r.h.s. of (\ref{label3.7}) the trace should be calculated over
spinorial indices of quark fields only. Making necessary contractions
we get
\begin{eqnarray}\label{label3.8} 
\hspace{-0.3in}&&{\rm tr}\{\langle \Omega |{\rm
T}(\eta^{(\alpha)}_{\Lambda^0}(x_1):\bar{u}(x + \frac{1}{2}\,y)\otimes
u(x - \frac{1}{2}\,y): \bar{\eta}^{(\beta)}_{\Lambda^0}(x_2))
|\Omega \rangle\} =\nonumber\\
\hspace{-0.3in}&& = 4{\rm tr}\{S^{(u)}_F(x - x_1 +
\frac{1}{2}\,y)\gamma^{\mu}S^{(s)}_F(x_1 -
x_2)\gamma^{\nu}S^{(u)}_F(x_2 - x +
\frac{1}{2}\,y)\}(\gamma_{\mu}\gamma^5S^{(d)}_F(x_1 -
x_2)\gamma_{\nu}\gamma^5)^{(\alpha\beta)}\nonumber\\
\hspace{-0.3in}&&+  4{\rm tr}\{\gamma^{\mu}S^{(s)}_F(x_1 -
x_2)\gamma^{\nu}S^{(d)}_F(x_2 -
x_1)\}(\gamma_{\mu}\gamma^5S^{(u)}_F(x_1 - x -
\frac{1}{2}\,y)S^{(u)}_F(x - x_2 -
\frac{1}{2}\,y)\gamma_{\nu}\gamma^5)^{(\alpha\beta)}\nonumber\\
\hspace{-0.3in}&& - 4(\gamma_{\mu}\gamma^5S^{(d)}_F(x_1 -
x_2)\gamma^{\nu}S^{(s)}_F(x_2 - x_1)\gamma^{\mu}S^{(u)}_F(x_1 - x -
\frac{1}{2}\,y)S^{(u)}_F(x - x_2 -
\frac{1}{2}\,y)\gamma_{\nu}\gamma^5)^{(\alpha\beta)}\nonumber\\
\hspace{-0.3in}&&- 4(\gamma_{\mu}\gamma^5S^{(u)}(x_1 - x -
\frac{1}{2}\,y)S^{(u)}(x - x_2 -
\frac{1}{2}\,y)\gamma^{\nu}S^{(s)}_F(x_2 - x_1)
\gamma^{\mu}S^{(d)}_F(x_1 - x_2)\gamma_{\nu}
\gamma^5)^{(\alpha\beta)}.\nonumber\\
\hspace{-0.3in}&&
\end{eqnarray}
In an analogous way one can calculate the matrix elements of the
operators $\bar{d}(x + \frac{1}{2}\,y)\otimes d(x - \frac{1}{2}\,y)$
and $\bar{s}(x + \frac{1}{2}\,y)\otimes s(x - \frac{1}{2}\,y)$. They
read
\begin{eqnarray}\label{label3.9} 
\hspace{-0.3in}&&{\rm tr}\{\langle \Omega|{\rm
T}(\eta^{(\alpha)}_{\Lambda^0}(x_1)\bar{d}(x + \frac{1}{2}\,y)\otimes
d(x - \frac{1}{2}\,y) \bar{\eta}^{(\beta)}_{\Lambda^0}(x_2))
|\Omega \rangle\} =\nonumber\\
\hspace{-0.3in}&&= 4{\rm tr}\{S^{(d)}_F(x - x_1 +
\frac{1}{2}\,y)\gamma^{\mu}S^{(s)}_F(x_1 -
x_2)\gamma^{\nu}S^{(d)}_F(x_2 - x +
\frac{1}{2}\,y)\}(\gamma_{\mu}\gamma^5S^{(u)}_F(x_1 -
x_2)\gamma_{\nu}\gamma^5)^{(\alpha\beta)}\nonumber\\
\hspace{-0.3in}&&+ 4{\rm tr}\{\gamma^{\mu}S^{(s)}_F(x_1 -
x_2)\gamma^{\nu}S^{(u)}_F(x_2 -
x_1)\}(\gamma_{\mu}\gamma^5S^{(d)}_F(x_1 - x -
\frac{1}{2}\,y)S^{(d)}_F(x - x_2 -
\frac{1}{2}\,y)\gamma_{\nu}\gamma^5)^{(\alpha\beta)} \nonumber\\
\hspace{-0.3in}&& - 4(\gamma_{\mu}\gamma^5S^{(u)}_F(x_1 -
x_2)\gamma^{\nu}S^{(s)}_F(x_2 - x_1)\gamma^{\mu}S^{(d)}_F(x_1 - x -
\frac{1}{2}\,y)S^{(d)}_F(x - x_2 -
\frac{1}{2}\,y)\gamma_{\nu}\gamma^5)^{(\alpha\beta)}\nonumber\\
\hspace{-0.3in}&&- 4(\gamma_{\mu}\gamma^5S^{(d)}(x_1 - x -
\frac{1}{2}\,y)S^{(d)}(x - x_2 -
\frac{1}{2}\,y)\gamma^{\nu}S^{(s)}_F(x_2 - x_1)
\gamma^{\mu}S^{(u)}_F(x_1 - x_2)\gamma_{\nu}
\gamma^5)^{(\alpha\beta)}\nonumber\\
\hspace{-0.3in}&&
\end{eqnarray}
and 
\begin{eqnarray}\label{label3.10} 
\hspace{-0.3in}&&{\rm tr}\{\langle \Omega |{\rm
T}(\eta^{(\alpha)}_{\Lambda^0}(x_1)\bar{s}(x + \frac{1}{2}\,y)\otimes
s(x - \frac{1}{2}\,y) \bar{\eta}^{(\beta)}_{\Lambda^0}(x_2)) |
\Omega \rangle\} =\nonumber\\
\hspace{-0.3in}&&= 4{\rm tr}\{S^{(u)}_F(x_2 -
x_1)\gamma^{\mu}S^{(s)}_F(x_1 - x + \frac{1}{2}\,y)S^{(s)}_F(x - x_2 +
\frac{1}{2}\,y)\gamma^{\nu}\}(\gamma_{\mu}\gamma^5S^{(d)}_F(x_1 -
x_2)\gamma_{\nu}\gamma^5)^{(\alpha\beta)}\nonumber\\
\hspace{-0.3in}&&+ 4{\rm tr}\{S^{(d)}_F(x_2 -
x_1)\gamma^{\mu}S^{(s)}_F(x_1 - x + \frac{1}{2}\,y)S^{(s)}_F(x - x_2 +
\frac{1}{2}\,y)\gamma^{\nu}\}(\gamma_{\mu}\gamma^5S^{(u)}_F(x_1 -
x_2)\gamma_{\nu}\gamma^5)^{(\alpha\beta)}\nonumber\\
\hspace{-0.3in}&&- 4(\gamma_{\mu}\gamma^5S^{(d)}_F(x_1 -
x_2)\gamma^{\nu}S^{(s)}_F(x_2 - x - \frac{1}{2}\,y)S^{(s)}_F(x - x_1 -
\frac{1}{2}\,y)\gamma^{\mu}S^{(u)}_F(x_1 -
x_2)\gamma_{\nu}\gamma^5)^{(\alpha\beta)}\nonumber\\
\hspace{-0.3in}&&- 4(\gamma_{\mu}\gamma^5S^{(u)}_F(x_1 -
x_2)\gamma^{\nu}S^{(s)}_F(x_2 - x - \frac{1}{2}\,y)S^{(s)}_F(x - x_1 -
\frac{1}{2}\,y)\gamma^{\mu}S^{(d)}_F(x_1 -
x_2)\gamma_{\nu}\gamma^5)^{(\alpha\beta)}.\nonumber\\
\hspace{-0.3in}&&
\end{eqnarray}
The two-point fermion Green function has the following form
\begin{eqnarray}\label{label3.11}
S_F(z)=\int \frac{d^4q}{(2\pi)^4}S_F(q)\,e^{\textstyle -iq\cdot
z}\, ,
\end{eqnarray}
where $S_F(q)$ is the Fourier transformation of the two-point Green
function $S_F(z)$. For finite temperature $T$ it is equal to
\cite{RJ74}
\begin{eqnarray}\label{label3.12}
\hspace{-0.3in}S_F(q) = - 2\pi\,i\,\theta(q^0)\, \delta(q^2 - m^2)
\,\frac {\displaystyle (\hat{q} + m)} {\displaystyle e^{\textstyle
(q\cdot U -\mu)/T} + 1},
\end{eqnarray}
where $U$ is a hydrodynamical velocity, $\mu$ is a quark chemical
potential of light $u$ and $d$ quarks related to the chemical
potential of the $\Lambda^0$ hyperon by $\mu_{\Lambda^0} = 2\mu$. The
chemical potential of strange quarks we set zero, since in our
approach they are massive with mass $m_s$.

For the non--equilibrium quark gas the quantities $U$, $\mu$ and $T$
should depend on a space--time point $x$ (\ref{label1.2}). The
hydrodynamical velocity $U(x)$, the quark chemical potential $\mu(x)$
and temperature $T(x)$ as functions of the space--time point $x$
should be obtained by solving transport equations within the
Quark--Gluon transport theory \cite{UH2}. However, as we show below
for the analysis of the polarization properties of $\Lambda^0$
hyperons produced from the QGP we do not need to know the explicit
expressions and $x$--dependence of the parameters $U(x)$, $\mu(x)$ and
$T(x)$.

We would like to emphasize that our definition of the two--point quark
Green function runs parallel to the ideology of the Parton Model
\cite{RF69}--\cite{FC79} and the Coalescence Quark Model \cite{TB83}.

Using the vacuum expectation values defined above we are able to
proceed to the momentum representation of the quark Green function and
to calculate the quantity
\begin{eqnarray}\label{label3.13}
\hspace{-0.5in}&&W^{(+)}(x,p; k)^{\alpha\beta} =\nonumber\\
\hspace{-0.5in}&&=\theta(p^0)\int d^4x_1 d^4x_2d^4y\,
e^{\textstyle\,ik\cdot (x_1 - x_2)}\,e^{\textstyle\,-ip\cdot y}\,{\rm
tr}\{\langle \Omega |{\rm
T}(\eta^{\alpha}_{\Lambda^0}(x_1)\hat{W}^{(+)}_u(x,
p)\,\bar{\eta}^{(\beta)}_{\Lambda^0}(x_2))|\Omega\rangle\nonumber\\
\hspace{-0.5in}&&+ \theta(p^0)\int d^4x_1
d^4x_2d^4y\,e^{\textstyle\,ik\cdot (x_1 -
x_2)}\,e^{\textstyle\,-ip\cdot y}\,{\rm tr}\{\langle \Omega |{\rm
T}(\eta^{\alpha}_{\Lambda^0}(x_1)\hat{W}^{(+)}_d(x,
p)\,\bar{\eta}^{(\beta)}_{\Lambda^0}(x_2))|\Omega\rangle\nonumber\\
\hspace{-0.5in}&&+ \theta(p^0)\int d^4x_1
d^4x_2d^4y\,e^{\textstyle\,ik\cdot (x_1 -
x_2)},e^{\textstyle\,-ip\cdot y}\,{\rm tr}\{\langle \Omega |{\rm
T}(\eta^{\alpha}_{\Lambda^0}(x_1)\hat{W}^{(+)}_s(x,
p)\,\bar{\eta}^{(\beta)}_{\Lambda^0}(x_2))|\Omega\rangle.\nonumber\\
\hspace{-0.5in}&&
\end{eqnarray}
We get
\begin{eqnarray}\label{label3.14}
&&\theta(p^0)\int d^4x_1 d^4x_2\,e^{\textstyle\,ik\cdot (x_1 -
x_2)}\,{\rm tr}\{\langle \Omega |{\rm
T}(\eta^{\alpha}_{\Lambda^0}(x_1)\hat{W}^{(+)}_u(x,
p)\,\bar{\eta}^{(\beta)}_{\Lambda^0}(x_2))|\Omega\rangle = \nonumber\\
&&= 4\,\theta(p^0)\int \frac{d^4q}{(2\pi)^4}\,{\rm tr}\{\gamma^{\mu}
S^{(s)}_F(- q) \gamma^{\nu} S^{(u)}_F( - p) S^{(u)}_F( -
p)\}\,(\gamma_{\mu}\gamma^5 S^{(d)}_F(q - p - k) \gamma_{\nu}
\gamma^5)^{(\alpha \beta)}\nonumber\\ &&+ 4\,\theta(p^0)\int
\frac{d^4q}{(2\pi)^4}\,{\rm tr}\{\gamma^{\mu} S^{(s)}_F(q)
\gamma^{\nu} S^{(d)}_F(q + p - k)\}\,(\gamma_{\mu}\gamma^5
S^{(u)}_F(p) S^{(u)}_F(p) \gamma_{\nu}\gamma^5)^{(\alpha
\beta)}\nonumber\\&&- 4\,\theta(p^0)\int
\frac{d^4q}{(2\pi)^4}\,(\gamma_{\mu} \gamma^5 S^{(d)}_F(q)
\gamma^{\nu} S^{(s)}_F(q + p - k) \gamma^{\mu}
S^{(u)}_F(p)S^{(u)}_F(p) \gamma_{\nu} \gamma^5)^{(\alpha\beta)}
\nonumber\\ &&- 4\,\theta(p^0)\int
\frac{d^4q}{(2\pi)^4}\,(\gamma_{\mu}\gamma^5 S^{(u)}_F(p)S^{(u)}_F(p)
\gamma^{\nu} S^{(s)}_F(q) \gamma^{\mu} S^{(d)}_F(q + k - p)
\gamma_{\nu} \gamma^5)^{(\alpha \beta)}
\end{eqnarray}
and 
\begin{eqnarray}\label{label3.15}
&&\theta(p^0)\int d^4x_1 d^4x_2\,e^{\textstyle\,ik\cdot (x_1 -
x_2)}\,{\rm tr}\{\langle \Omega |{\rm
T}(\eta^{\alpha}_{\Lambda^0}(x_1)\hat{W}^{(+)}_d(x,
p)\,\bar{\eta}^{(\beta)}_{\Lambda^0}(x_2))|\Omega\rangle = \nonumber\\
&&= 4\,\theta(p^0)\int \frac{d^4q}{(2\pi)^4}\,(\gamma_{\mu} \gamma^5
S^{(u)}_F(q) \gamma^{\nu} S^{(s)}_F(q + p - k) \gamma^{\mu}
S^{(d)}_F(p)S^{(d)}_F(p) \gamma_{\nu} \gamma^5)^{(\alpha\beta)}
\nonumber\\ &&+ 4\,\theta(p^0)\int
\frac{d^4q}{(2\pi)^4}\,(\gamma_{\mu}\gamma^5 S^{(d)}_F(p)S^{(d)}_F(p)
\gamma^{\nu} S^{(s)}_F(q) \gamma^{\mu} S^{(u)}_F(q + k - p)
\gamma_{\nu} \gamma^5)^{(\alpha \beta)}\nonumber\\ &&-
4\,\theta(p^0)\int \frac{d^4q}{(2\pi)^4}\,{\rm tr}\{\gamma^{\mu}
S^{(s)}_F(- q) \gamma^{\nu} S^{(d)}_F( - p) S^{(d)}_F( -
p)\}\,(\gamma_{\mu}\gamma^5 S^{(u)}_F(q - p - k) \gamma_{\nu}
\gamma^5)^{(\alpha \beta)}\nonumber\\ &&- 4\,\theta(p^0)\int
\frac{d^4q}{(2\pi)^4}\,{\rm tr}\{\gamma^{\mu} S^{(s)}_F(q)
\gamma^{\nu} S^{(u)}_F(q + p - k)\}\,(\gamma_{\mu}\gamma^5
S^{(d)}_F(p) S^{(d)}_F(p) \gamma_{\nu}\gamma^5)^{(\alpha
\beta)}\nonumber\\
\end{eqnarray}
and 
\begin{eqnarray}\label{label3.16}
\hspace{-0.5in}&&\theta(p^0)\int d^4x_1 d^4x_2\,e^{\textstyle\,ik\cdot
(x_1 - x_2)}\,{\rm tr}\{\langle \Omega |{\rm
T}(\eta^{\alpha}_{\Lambda^0}(x_1)\hat{W}^{(+)}_s(x,
p)\,\bar{\eta}^{(\beta)}_{\Lambda^0}(x_2))|\Omega\rangle = \nonumber\\
\hspace{-0.5in}&&= 4\,\theta(p^0)\int
\frac{d^4q}{(2\pi)^4}\,(\gamma_{\mu}\gamma^5 S^{(d)}_F(q) \gamma^{\nu}
S^{(s)}_F(p)S^{(s)}_F(p) \gamma^{\mu} S^{(u)}_F(p + k - q)
\gamma_{\nu} \gamma^5)^{(\alpha \beta)}\nonumber\\ \hspace{-0.5in}&&+
4\,\theta(p^0)\int \frac{d^4q}{(2\pi)^4}\,(\gamma_{\mu}\gamma^5
S^{(u)}_F(q) \gamma^{\nu} S^{(s)}_F(p)S^{(s)}_F(p) \gamma^{\mu}
S^{(d)}_F(p + k - q) \gamma_{\nu} \gamma^5)^{(\alpha
\beta)}\nonumber\\ \hspace{-0.5in}&&- 4\,\theta(p^0)\int
\frac{d^4q}{(2\pi)^4}\,{\rm tr}\{S^{(u)}_F(q) \gamma^{\mu}
S^{(s)}_F(-p) S^{(s)}_F(-p) \gamma^{\nu}\}\,(\gamma_{\mu} \gamma^5
S^{(d)}_F(q + p + k) \gamma_{\nu} \gamma^5)^{(\alpha
\beta)}\nonumber\\ \hspace{-0.5in}&&- 4\,\theta(p^0)\int
\frac{d^4q}{(2\pi)^4}\,{\rm tr}\{S^{(d)}_F(q) \gamma^{\mu}
S^{(s)}_F(-p) S^{(s)}_F(-p) \gamma^{\nu}\}\,(\gamma_{\mu} \gamma^5
S^{(u)}_F(q + p + k) \gamma_{\nu} \gamma^5)^{(\alpha \beta)}.
\end{eqnarray}
Since the quarks $u$ and $d$ are massless and the strange quark $s$ is
massive with mass $m_s$, we can set
\begin{eqnarray}\label{label3.17}
S^{(u)}_F(p)S^{(u)}_F(p) = S^{(d)}_F(p)S^{(d)}_F(p) = 0.
\end{eqnarray}
Indeed, using (\ref{label3.12}) we get
\begin{eqnarray}\label{label3.18}
S^{(u)}_F(p)S^{(u)}_F(p) &=& - (2\pi)^2
\delta(p^2)\frac{\hat{p}}{\displaystyle e^{\textstyle\,(p\cdot U -
\mu)/T} + 1}\delta(p^2)\frac{\hat{p}}{\displaystyle
e^{\textstyle\,(p\cdot U - \mu)/T} + 1} =\nonumber\\ &=& - (2\pi)^2
\delta(0)\,\frac{p^2\delta(p^2)}{\displaystyle
\Big(e^{\textstyle\,(p\cdot U - \mu)/T} + 1\Big)^2} = 0,
\end{eqnarray}
where we have used the property of the $\delta$--function
$p^2\delta(p^2) = 0$\,\footnote{The infinity related to $\delta(0)$
can be deleted by introducing the rate of the number of $\Lambda^0$
hyperons (see Eq.(\ref{label4.12}) in Section 4).}.

As a result the quantity $W^{(+)}(x,p; k)^{\alpha\beta}$ is defined by
the contribution of the strange $s$--quark only
\begin{eqnarray}\label{label3.19}
\hspace{-0.5in}&&W^{(+)}(x,p; k)^{\alpha\beta} =\nonumber\\
\hspace{-0.5in}&&= 4\,\theta(p^0)\int
\frac{d^4q}{(2\pi)^4}\,{\rm tr}\{S^{(u)}_F(q) \gamma^{\mu}
S^{(s)}_F(-p) S^{(s)}_F(-p) \gamma^{\nu}\}\,(\gamma_{\mu} \gamma^5
S^{(d)}_F(q + p + k) \gamma_{\nu} \gamma^5)^{(\alpha
\beta)}\nonumber\\ \hspace{-0.5in}&&+ 4\,\theta(p^0)\int
\frac{d^4q}{(2\pi)^4}\,{\rm tr}\{S^{(d)}_F(q) \gamma^{\mu}
S^{(s)}_F(-p) S^{(s)}_F(-p) \gamma^{\nu}\}\,(\gamma_{\mu} \gamma^5
S^{(u)}_F(q + p + k) \gamma_{\nu} \gamma^5)^{(\alpha \beta)}\nonumber\\
\hspace{-0.5in}&&- 4\,\theta(p^0)\int
\frac{d^4q}{(2\pi)^4}\,(\gamma_{\mu}\gamma^5 S^{(d)}_F(q) \gamma^{\nu}
S^{(s)}_F(p)S^{(s)}_F(p) \gamma^{\mu} S^{(u)}_F(p + k - q)
\gamma_{\nu} \gamma^5)^{(\alpha \beta)}\nonumber\\ \hspace{-0.5in}&&-
4\,\theta(p^0)\int \frac{d^4q}{(2\pi)^4}\,(\gamma_{\mu}\gamma^5
S^{(u)}_F(q) \gamma^{\nu} S^{(s)}_F(p)S^{(s)}_F(p) \gamma^{\mu}
S^{(d)}_F(p + k - q) \gamma_{\nu} \gamma^5)^{(\alpha \beta)}.
\end{eqnarray}
Thus, in our approach the polarization properties of the $\Lambda^0$
hyperon are fully defined by the strange $s$--quark. This agrees well
with the DeGrand--Miettinen model of polarization of $\Lambda^0$
hyperons \cite{TD81} and the experimental data on polarization of
$\Lambda^0$ hyperons in proton--induced reactions testifying that the
spin of the $\Lambda^0$ hyperon is caused by the strange $s$--quark
and that the $u$ and $d$ quarks are combined into a diquark with zero
total angular momentum and isospin \cite{BB88}.

Due to the Heaviside function $\theta(p^0)$ the r.h.s. of
(\ref{label3.19}) can be transcribed into the form
\begin{eqnarray}\label{label3.20}
\hspace{-0.3in}&&W^{(+)}(x,p; k)^{\alpha\beta} = \nonumber\\
\hspace{-0.3in}&&= -\,8\,\theta(p^0)\int
\frac{d^4q}{(2\pi)^4}(\gamma_{\mu}\gamma^5 S^{(d)}_F(q) \gamma^{\nu}
S^{(s)}_F(p)S^{(s)}_F(p) \gamma^{\mu} S^{(u)}_F(p + k - q)
\gamma_{\nu} \gamma^5)^{(\alpha \beta)}.
\end{eqnarray}
Now we are able to calculate the momentum distribution of $\Lambda^0$
hyperons produced from the QGP in dependence on the
$\Lambda^0$--hyperon polarization.

\section{Polarization of $\Lambda^0$ hyperons}
\setcounter{equation}{0}

Substituting (\ref{label3.20}) in (\ref{label2.1}) we obtain the
momentum distribution of $\Lambda^0$ hyperons produced from the QGP in
terms of the Green functions of quarks treated in the non--equilibrium
state with $U$, $\mu$ and $T$ depending on a space--time point $x$. It
reads
\begin{eqnarray}\label{label4.1}
\hspace{-0.3in}E_{\vec{k}}\,\frac{d^3 N_{\Lambda^0}(\vec{k}\,)}{d^3k}
&=& \frac{g^2_B}{m_{\Lambda^0}}\int_{\Sigma} d\sigma^{\mu}(x)
k_{\mu}\,\theta(k^0) \int \frac{d^4p d^4q}{(2\pi)^4}\,(- \,1)\,{\rm
tr}\{\gamma_{\mu}\gamma^5 S^{(d)}_F(x; q)\gamma^{\nu}
\nonumber\\
\hspace{-0.3in}&\times&S^{(s)}_F(x; p)S^{(s)}_F(x; p) \gamma^{\mu}
S^{(u)}_F(x; p + k - q) \gamma_{\nu} \gamma^5 (\hat{k} +
m_{\Lambda^0})(1 + \gamma^5 \hat{\zeta})\}.
\end{eqnarray}
Using the expressions for the quark Green functions (\ref{label3.12})
with $U$, $\mu$ and $T$, replaced by $U(x)$, $\mu(x)$ and $T(x)$, we
reduce the r.h.s. of (\ref{label4.1}) to the form
\begin{eqnarray}\label{label4.2}
\hspace{-0.3in}&&E_{\vec{k}}\,\frac{d^3
N_{\Lambda^0}(\vec{k}\,)}{d^3k} = 2\,g^2_B\,
\frac{m_s}{m_{\Lambda^0}}\int_{\Sigma} d\sigma^{\mu}(x)
k_{\mu}\,\theta(k^0) \int d^4p d^4q\,\theta(p^0)\,\frac{\delta(p^2 -
m^2_s)}{\displaystyle e^{\textstyle\,p\cdot U(x)/T(x)} + 1}\nonumber\\
\hspace{-0.3in}&&\times\,\frac{\delta(p^2 - m^2_s)}{\displaystyle
e^{\textstyle\,p\cdot U(x)/T(x)} +
1}\,\frac{\theta(q^0)\delta(q^2)}{\displaystyle e^{\textstyle\,(q\cdot
U(x) - \mu(x))/T(x)} + 1}\nonumber\\
\hspace{-0.3in}&&\times\,\frac{\theta(p^0 + k^0 - q^0)\delta((p + k -
q)^2)}{\displaystyle e^{\textstyle\,((p + k - q)\cdot U(x) -
\mu(x))/T(x)} + 1}\nonumber\\ \hspace{-0.3in}&&\times\,(-\,1)\,{\rm
tr}\{\gamma_{\mu}\gamma^5 \hat{q} \gamma^{\nu}(\hat{p} + m_s)
\gamma^{\mu}(\hat{p} + \hat{k} - \hat{q})\gamma_{\nu} \gamma^5
(\hat{k} + m_{\Lambda^0})(1 + \gamma^5 \hat{\zeta})\}.
\end{eqnarray}
The trace over $\gamma$--matrices is equal to
\begin{eqnarray}\label{label4.3}
&&(-1)\,{\rm tr}\{\gamma_{\mu}\gamma^5 \hat{q} \gamma^{\nu}(\hat{p} +
m_s) \gamma^{\mu}(\hat{p} + \hat{k} - \hat{q})\gamma_{\nu} \gamma^5
(\hat{k} + m_{\Lambda^0})(1 + \gamma^5 \hat{\zeta})\} =\nonumber\\ &&=
(16(k\cdot p) + 8 m_s m_{\Lambda^0})(p + k)^2 + 16im_s
\varepsilon^{\mu\nu\alpha\beta}p_{\mu}q_{\nu}k_{\alpha}\zeta_{\beta},
\end{eqnarray}
where we have taken into account the properties of the
$\delta$--functions $\delta(q^2)$ and $\delta((p + k - q)^2)$.  The
r.h.s. of (\ref{label4.2}) we suggest to rewrite as follows
\begin{eqnarray}\label{label4.4}
\hspace{-0.3in}&&E_{\vec{k}}\,\frac{d^3
N_{\Lambda^0}(\vec{k}\,)}{d^3k} = \delta(0)\,g^2_B\,
\frac{m_s}{m_{\Lambda^0}}\int_{\Sigma} d\sigma^{\mu}(x)
k_{\mu}\,\theta(k^0) \int \frac{d^4p d^4q}{\sqrt{\vec{p}^{\,2} +
m^2_s}}\,\frac{\theta(p^0)\delta(p^2 - m^2_s)}{\displaystyle
\Big(e^{\textstyle\,p\cdot U(x)/T(x)} + 1\Big)^2}\nonumber\\
\hspace{-0.3in}&&\times\,\frac{\theta(q^0)\delta(q^2)}{\displaystyle
e^{\textstyle\,(q\cdot U(x) - \mu(x))/T(x)} + 1}\,\frac{\theta(p^0 +
k^0 - q^0)\delta((p + k - q)^2)}{\displaystyle e^{\textstyle\,((p + k
- q)\cdot U(x) - \mu(x))/T(x)} + 1}\nonumber\\
\hspace{-0.3in}&&\times\,(-\,1)\,{\rm tr}\{\gamma_{\mu}\gamma^5
\hat{q} \gamma^{\nu}(\hat{p} + m_s) \gamma^{\mu}(\hat{p} + \hat{k} -
\hat{q})\gamma_{\nu} \gamma^5 (\hat{k} + m_{\Lambda^0})(1 + \gamma^5
\hat{\zeta})\},
\end{eqnarray}
where we have used the definition of the $\delta$--function
\begin{eqnarray}\label{label4.5}
\theta(p^0)\,\delta(p^2 - m^2_s) = \theta(p^0)\,\frac{\delta(p^0 -
\sqrt{\vec{p}^{\,2} + m^2_s})}{2\sqrt{\vec{p}^{\,2} + m^2_s}}
\end{eqnarray}
According to this definition the $\delta$--function $\delta(0)$ can be
determined as
\begin{eqnarray}\label{label4.6}
\delta(0) = \lim_{\tau \to \infty} \frac{\tau}{2\pi}.
\end{eqnarray}
Substituting (\ref{label4.6}) into (\ref{label4.4}), dividing both sides
by $\tau$ and introducing the notation
\begin{eqnarray}\label{label4.7}
\dot{N}_{\Lambda^0}(\vec{k}\,) = \lim_{\tau \to \infty}
\frac{N_{\Lambda^0}(\vec{k}\,)}{\tau}
\end{eqnarray}
for the rate of $\Lambda^0$ hyperons produced from  the QGP we get
\begin{eqnarray}\label{label4.8}
\hspace{-0.3in}&&E_{\vec{k}}\,\frac{d^3
\dot{N}_{\Lambda^0}(\vec{k}\,)}{d^3k} = \frac{g^2_B}{2\pi}\,
\frac{m_s}{m_{\Lambda^0}}\int_{\Sigma} d\sigma^{\mu}(x)
k_{\mu}\,\theta(k^0) \int \frac{d^4p d^4q}{\sqrt{\vec{p}^{\,2} +
m^2_s}}\,\frac{\theta(p^0)\delta(p^2 - m^2_s)}{\displaystyle
\Big(e^{\textstyle\,p\cdot U(x)/T(x)} + 1\Big)^2}\nonumber\\
\hspace{-0.3in}&&\times\,\frac{\theta(q^0)\delta(q^2)}{\displaystyle
e^{\textstyle\,(q\cdot U(x) - \mu(x))/T(x)} + 1}\frac{\theta(p^0 + k^0
- q^0)\delta((p + k - q)^2)}{\displaystyle e^{\textstyle\,((p + k -
q)\cdot U(x) - \mu(x))/T(x)} + 1}\nonumber\\
\hspace{-0.3in}&&\times\,(-\,1)\,{\rm tr}\{\gamma_{\mu}\gamma^5
\hat{q} \gamma^{\nu}(\hat{p} + m_s) \gamma^{\mu}(\hat{p} + \hat{k} -
\hat{q})\gamma_{\nu} \gamma^5 (\hat{k} + m_{\Lambda^0})(1 + \gamma^5
\hat{\zeta})\}.
\end{eqnarray}
Substituting (\ref{label4.3}) in (\ref{label4.8}) we end up with the
expression
\begin{eqnarray}\label{label4.9}
\hspace{-0.3in}&&E_{\vec{k}}\,\frac{d^3
\dot{N}_{\Lambda^0}(\vec{k}\,)}{d^3k} = 32\,\frac{g^2_B}{2\pi}\,
\frac{m_s}{m_{\Lambda^0}}\int_{\Sigma} d\sigma^{\mu}(x)
k_{\mu}\,\theta(k^0) \int \frac{d^4p d^4q}{\sqrt{\vec{p}^{\,2} +
m^2_s}}\,\frac{\theta(p^0)\delta(p^2 - m^2_s)}{\displaystyle
\Big(e^{\textstyle\,p\cdot U(x)/T(x)} + 1\Big)^2}\nonumber\\
\hspace{-0.3in}&&\times\,\frac{\theta(q^0)\delta(q^2)}{\displaystyle
e^{\textstyle\,(q\cdot U(x) - \mu(x))/T(x)} + 1}\frac{\theta(p^0 + k^0
- q^0)\delta((p + k - q)^2)}{\displaystyle e^{\textstyle\,((p + k -
q)\cdot U(x) - \mu(x))/T(x)} + 1}\nonumber\\ \hspace{-0.3in}&&\times\,\Big[\Big((k\cdot
p) + \frac{1}{2}\,m_s m_{\Lambda^0}\Big)(p + k)^2 + im_s
\varepsilon^{\mu\nu\alpha\beta} p_{\mu} q_{\nu} k_{\alpha}
\zeta_{\beta}\Big].
\end{eqnarray}
It is obvious that integration over the momenta $p$ and $q$ leads to
the vanishing of the term proportional to the polarization vector of
$\Lambda^0$ hyperons $\zeta$.  This means that $\Lambda^0$ hyperons
produced from the QGP are unpolarized.

Now let us show that the momentum distribution (\ref{label4.9}) can be
reduces to the form analogous to (\ref{label1.2}).  For this aim we
assume that the Fermi--Dirac gasses of quarks and $\Lambda^0$ hyperons
can be described well in the Boltzmann gas approximation
\cite{UH1}. This allows to neglect the contributions of unities with
respect to exponentials in the distribution functions of quarks and
$\Lambda^0$ hyperons.

In the Boltzmann quark gas approximation the integrand of
(\ref{label4.9}) can be transcribed into the form
\begin{eqnarray}\label{label4.10}
\hspace{-0.3in}&&E_{\vec{k}}\,\frac{d^3
\dot{N}_{\Lambda^0}(\vec{k}\,)}{d^3k} = 32\,\frac{g^2_B}{2\pi}\,
\frac{m_s}{m_{\Lambda^0}}\int_{\Sigma} d\sigma^{\mu}(x)
k_{\mu}\,\theta(k^0)\,e^{\textstyle\,-\,(k\cdot U(x) -
\mu_{\Lambda^0}(x))/T(x)}\nonumber\\
\hspace{-0.3in}&&\times\int \frac{d^4p}{\sqrt{\vec{p}^{\,2} +
m^2_s}}\,\theta(p^0)\,\delta(p^2 - m^2_s)\,\Big((k\cdot p) +
\frac{1}{2}\,m_s m_{\Lambda^0}\Big)\,(p + k)^2\,e^{\textstyle\,-
3p\cdot U(x)/T(x)}\nonumber\\
\hspace{-0.3in}&&\times\int d^4q \,\theta(q^0)\delta(q^2)\,\theta(p^0
+ k^0 - q^0)\delta((p + k - q)^2),
\end{eqnarray}
where we have used $\mu_{\Lambda^0}(x) = 2\mu(x)$. Then, the function
\begin{eqnarray}\label{label4.11}
e^{\textstyle\,-\,(k\cdot U(x) - \mu_{\Lambda^0}(x))/T(x)}
\end{eqnarray}
coincides with the distribution function of the $\Lambda^0$ hyperon
(\ref{label1.2}) in the Boltzmann gas approximation. This testifies
the correctness of our quark level approach to the description of the
momentum distribution of the number of $\Lambda^0$ hyperons produced
from the QGP. The calculation of the integrals over $q$ and $p$ in
(\ref{label4.10}) are adduced in the Appendix A and B.

Using the results obtained in Appendix A and B we get
\begin{eqnarray}\label{label4.12}
E_{\vec{k}}\,\frac{d^3 \dot{N}_{\Lambda^0}(\vec{k}\,)}{d^3k} =
\frac{2}{(2\pi)^3}\int_{\Sigma} d\sigma^{\mu}(x)
k_{\mu}\,\theta(k^0)\,F(U(x),T(x),k)\,f_{\Lambda^0}(x, k).
\end{eqnarray}
We have denoted $F(U(x),T(x),k) = 4\pi^3\,I(U(x),T(x),k)$ (see (B.7)
of the Appendix B).

In our approach Eq.(\ref{label4.12}) describes the momentum
distribution of the rate of $\Lambda^0$ hyperons produced from the
QGP. Since it does not depend on the polarization of $\Lambda^0$
hyperons, in our approach $\Lambda^0$ hyperons, produced from the QGP,
are unpolarized. This agrees well with the results obtained within
other theoretical approaches \cite{RS82}--\cite{MA01,AA02}.

\section{Conclusion}

We have analysed the polarization properties of $\Lambda^0$ hyperons
produced from the QGP. We have described the momentum distribution of
$\Lambda^0$ hyperons, produced from the QGP, in terms of the matrix
elements of the relativistic quark Wigner operators. The matrix
elements of these operators we have calculated within the Effective
quark model with chiral $U(3)\times U(3)$ symmetry and the
Quark--Gluon transport theory.  We have shown that using the quark
distribution functions in the form of the J\"uttner distribution
functions with a hydrodynamical 4--velocity $U(x)$, a quark chemical
potential $\mu(x)$ and a temperature $T(x)$ depending on the
space--time point $x$, the momentum distribution of $\Lambda^0$
hyperons can be represented in the form of the integral over the
freeze--out surface of the J\"utther distribution function of
$\Lambda^0$ hyperons. We have shown that without solving the
Quark--Gluon transport equations for the parameters $U(x)$, $\mu(x)$
and $T(x)$ the momentum distribution of the rate of $\Lambda^0$
hyperons does not depend of the polarization of $\Lambda^0$
hyperons. This means that $\Lambda^0$ hyperons are unpolarized, when
they are produced from the QGP.

Since it is well--known that in high--energy nuclear reactions
$\Lambda^0$ hyperons are produced highly polarized, the obtained
depolarization of the $\Lambda^0$ hyperons in the QGP can serve as a
one more signature of the QGP.

We would like to emphasize that in our approach the momentum
distribution of the rate of $\Lambda^0$ hyperons has turned out to be
defined by the matrix elements of the relativistic strange quark
Wigner operator only.  Thus, in our approach the polarization
properties of $\Lambda^0$ hyperons are fully determined by the spin of
the strange $s$--quarks. This agrees well with the DeGrand--Miettinen
model and the experimental data on the $\Lambda^0$ hyperon production
in proton--induced nuclear reactions \cite{BB88}. Such an agreement
testifies the correctness of the application of the relativistic quark
Wigner operators and the Effective quark model with chiral $U(3)\times
U(3)$ symmetry to the analysis of baryon production by the QGP in
ultra--relativistic heavy--ion collisions.

Our result concerning the production of unpolarized $\Lambda^0$
hyperons from the QGP agrees well with the results obtained within
other theoretical approaches \cite{RS82}--\cite{MA01,AA02}.

\newpage

\section*{Appendix A. Calculation of the integral over $q$}

Let us denote the integral over $q$ as $I(p,k)$. The calculation of
$I(p,k)$ runs in the way
$$
I(p,k) = \int d^4q \,\theta(q^0)\delta(q^2)\,\theta(p^0 + k^0 -
q^0)\delta((p + k - q)^2) = 
$$
$$
= \int d^3q\int^{+\infty}_{-\infty}dq^0\,\,
\frac{\theta(q^0)}{2|\vec{q}\,|}\,\delta(q^0 -
|\vec{q}\,|)\,\frac{\theta(p^0 + k^0 - q^0)}{2|\vec{p} + \vec{k} -
\vec{q}\,|}\,\delta(p^0 + k^0 - q^0 - |\vec{p} + \vec{k} -
\vec{q}\,|).\eqno(A.1)
$$
Integrating over $q^0$ we obtain
$$
I(p,k) = \int \frac{d^3q}{2|\vec{q}\,|}\,\frac{\theta(p^0 + k^0 -
|\vec{q}\,|)}{2|\vec{p} + \vec{k} - \vec{q}\,|}\,\delta(p^0 + k^0 -
 |\vec{q}\,| - |\vec{p} + \vec{k} - \vec{q}\,|).\eqno(A.2)
$$
In the spherical coordinates the integral over $\vec{q}$ reads
$$
I(p,k) = \frac{1}{4}\int^{p^0 + k^0}_0 d|\vec{q}\,||\vec{q}\,|\int
\frac{d\Omega_{\vec{q}}}{|\vec{p} + \vec{k} - \vec{q}\,|}\,\delta(p^0
+ k^0 - |\vec{q}\,| - |\vec{p} + \vec{k} - \vec{q}\,|),\eqno(A.3)
$$
where $d\Omega_{\vec{q}}$ is the solid angle.

The integration over the solid angle runs as follows
$$
\int \frac{d\Omega_{\vec{q}}}{|\vec{p} + \vec{k} -
\vec{q}\,|}\,\delta(p^0 + k^0 - |\vec{q}\,| - |\vec{p} + \vec{k} -
\vec{q}\,|) = 2\pi\int^{\pi}_0
\frac{d\theta\,\sin\theta}{\sqrt{(\vec{p} + \vec{k}\,)^{\,2} +
\vec{q}^{\,2} - 2|\vec{p} + \vec{k}\,||\vec{q}\,|\cos\theta}}
$$
$$
\times\,\delta(p^0 + k^0 - |\vec{q}\,| - \sqrt{(\vec{p} +
\vec{k}\,)^{\,2} + \vec{q}^{\,2} - 2|\vec{p} +
\vec{k}\,||\vec{q}\,|\cos\theta}\,).\eqno(A.4)
$$
Making a change of variables
$$
t = \sqrt{(\vec{p} +
\vec{k}\,)^{\,2} + \vec{q}^{\,2} - 2|\vec{p} +
\vec{k}\,||\vec{q}\,|\cos\theta}\eqno(A.5)
$$
we get
$$
\int \frac{d\Omega_{\vec{q}}}{|\vec{p} + \vec{k} -
\vec{q}\,|}\,\delta(p^0 + k^0 - |\vec{q}\,| - |\vec{p} + \vec{k} -
\vec{q}\,|) = 
$$
$$
= \frac{2\pi}{|\vec{q}\,||\vec{p} + \vec{k}\,|}\int^{|(\vec{p} +
\vec{k}\,|+ |\vec{q}\,|}_{||(\vec{p} + \vec{k}\,|-
|\vec{q}\,||}dt\,\delta(p^0 + k^0 - |\vec{q}\,| - t) =
$$
$$
= \frac{2\pi}{|\vec{q}\,||\vec{p} + \vec{k}\,|}\,\Big[\theta(p^0 +
k^0 - |\vec{q}\,| - ||\vec{p} + \vec{k}\,|- |\vec{q}\,||) -
\theta(p^0 + k^0 - |\vec{p} + \vec{k}\,| -
2|\vec{q}\,|)\Big].\eqno(A.6)
$$
Thus, $I(p,k)$ is defined by the expression
$$
I(p,k) = \frac{\pi}{2|\vec{p} + \vec{k}\,|}\int^{p^0 + k^0}_0
d|\vec{q}\,|\Big[\theta(p^0 + k^0 - |\vec{q}\,| - ||\vec{p} +
\vec{k}\,|- |\vec{q}\,||) - \theta(p^0 + k^0 - |\vec{p} + \vec{k}\,|
- 2|\vec{q}\,|)\Big].\eqno(A.7)
$$
Integrating by parts we get
$$
I(p,k) = \frac{\pi}{2|\vec{p} + \vec{k}\,|}\int^{p^0 + k^0}_0
d|\vec{q}\,||\vec{q}\,|\Big[\theta(|\vec{q}\,|- |\vec{p} + \vec{k}\,|)
\delta\Big(\frac{p^0 + k^0 + |\vec{p} + \vec{k}\,|}{2} -
|\vec{q}\,|\Big)
$$
$$
 - \delta\Big(\frac{(p^0 + k^0 - |\vec{p} + \vec{k}\,|}{2} -
|\vec{q}\,|\Big)\Big] = \frac{\pi}{2}.\eqno(A.8)
$$
Hence, the integral over $p$ does not depend on the integral over $q$.

\section*{Appendix B. Calculation of the integral over $p$}

Let us denote the integrals over $q$ and $p$ as $I(U,T,k)$
$$
I(U,T,k) = 8\,g^2_B\, \frac{m_s}{m_{\Lambda^0}}\int
\frac{d^4p}{\sqrt{\vec{p}^{\;2} + m^2_s}}\,\theta(p^0)\,\delta(p^2 -
m^2_s)
$$
$$
\times\,\Big((k\cdot p) + \frac{1}{2}\,m_s m_{\Lambda^0}\Big)\,(p +
k)^2\,e^{\textstyle\,- 3\,p\cdot U(x)/T(x)}.\eqno(B.1)
$$
Due to the Heaviside function $\theta(p^0)$ we can rewrite the
integral as follows
$$
I(U,T,k) = 4\,g^2_B\, \frac{m_s}{m_{\Lambda^0}}\int
d^3p\int^{\infty}_0\frac{dp^0}{\vec{p}^{\,2} + m^2_s}\,\delta(p^0 -
\sqrt{\vec{p}^{\,2} + m^2_s})\Big(k^0p^0 - \vec{k}\cdot \vec{p} +
\frac{1}{2}\,m_s m_{\Lambda^0}\Big)
$$
$$
\times\,(m^2_{\Lambda^0} + m^2_s + 2k^0p^0 - 2\vec{k}\cdot
\vec{p}\,)\,\exp\,\Big\{- 3\,p^0\,\frac{U^0(x)}{T(x)} +
3\,\vec{p}\cdot \frac{\vec{U}(x)}{T(x)}\Big\}.\eqno(B.2)
$$
Integrating over $p^0$ we get
$$
I(U,T,k) = 4\,g^2_B\, \frac{m_s}{m_{\Lambda^0}}\int
\frac{d^3p}{\vec{p}^{\,2} + m^2_s}\,\Big(k^0\sqrt{\vec{p}^{\,2} +
m^2_s} - \vec{k}\cdot \vec{p} + \frac{1}{2}\,m_s m_{\Lambda^0}\Big)
$$
$$
\times\,(m^2_{\Lambda^0} + m^2_s + 2k^0\sqrt{\vec{p}^{\,2} + m^2_s} -
2\vec{k}\cdot \vec{p}\,)\,\exp\,\Big\{- 3\,\sqrt{\vec{p}^{\,2} +
m^2_s}\,\frac{U^0(x)}{T(x)} + 3\,\vec{p}\cdot
\frac{\vec{U}(x)}{T(x)}\,\Big\}.\eqno(B.3)
$$
Introducing the notations $\rho^0 = U^0(x)/T(x)$ and $\vec{\rho} =
\vec{U}(x)/T(x)$ we can transcribe the r.h.s. of (B.3) as follows
$$
I(U,T,k) = g^2_B\, \frac{4}{9}\,\frac{m_s}{m_{\Lambda^0}}\,\Big(-
k^0\frac{\partial }{\partial \rho^0} - \vec{k}\cdot
\frac{\partial}{\partial \vec{\rho}} + \frac{1}{2}\,m_s
m_{\Lambda^0}\Big)
$$
$$
\times\,\Big(m^2_{\Lambda^0} + m^2_s - 2k^0\frac{\partial }{\partial
\rho^0} - 2\vec{k}\cdot \frac{\partial}{\partial \vec{\rho}}\Big)\int
\frac{d^3p}{\vec{p}^{\,2} + m^2_s}\,e^{\textstyle\,-
3\,\sqrt{\vec{p}^{\,2} + m^2_s}\,\rho^0 + 3\,\vec{p}\cdot
\vec{\rho}}.\eqno(B.4)
$$
The integral over $\vec{p}$ can be calculated as follows
$$
\int \frac{d^3p}{\vec{p}^{\,2} + m^2_s}\,e^{\textstyle\,-
3\,\sqrt{\vec{p}^{\,2} + m^2_s}\,\rho^0 + 3\,\vec{p}\cdot \vec{\rho}}
= \frac{4\pi}{3|\vec{\rho}\,|}\int^{\infty}_0\frac{dp
p\sinh(3p|\vec{\rho}\,|)}{p^2 + m^2_s}\,e^{\textstyle\,-
3\,\sqrt{p^2 + m^2_s}\,\rho^0},\eqno(B.5)
$$
where we have denoted $p = |\vec{p}\,|$. The integral over $p$ is
convergent. However, it is rather complicated to be expressed in terms
of special functions. Therefore, we leave this integral in the form
given by (B.5).

Thus, the result of the integration over $q$ and $p$ can be written as
$$
I(U,T,k) = g^2_B\, \frac{16\pi}{27}\,\frac{m_s}{m_{\Lambda^0}}\,\Big(-
k^0\frac{\partial }{\partial \rho^0} - \vec{k}\cdot
\frac{\partial}{\partial \vec{\rho}} + \frac{1}{2}\,m_s
m_{\Lambda^0}\Big)
$$
$$
\times\,\Big(m^2_{\Lambda^0} + m^2_s - 2k^0\frac{\partial }{\partial
\rho^0} - 2\vec{k}\cdot \frac{\partial}{\partial
\vec{\rho}}\Big)\frac{1}{|\vec{\rho}\,|}\int^{\infty}_0\frac{dp
p\sinh(p|\vec{\rho}\,|)}{p^2 + 9 m^2_s}\,e^{\textstyle\,- \sqrt{p^2 +
9m^2_s}\,\rho^0},\eqno(B.7)
$$
where we have changed the scale of the momentum $p \to p/3$.

\end{document}